# Dielectric tensor of perovskite oxides at finite temperature using equivariant graph neural network potentials

Alex Kutana, Koki Yoshimochi and Ryoji Asahi✉

Nagoya University, Furo-cho, Chikusa-ku, Nagoya, Japan
✉ E-mail: asahi.ryoji.d9@f.mail.nagoya-u.ac.jp

## Abstract
Atomistic simulations of properties of materials at finite temperatures are computationally demanding and require models that are more efficient than the ab initio approaches. Machine learning (ML) and artificial intelligence (AI) address this issue by enabling accurate models with close to ab initio accuracy. Here, we demonstrate the utility of ML models in capturing properties of realistic materials by performing finite temperature molecular dynamics simulations of perovskite oxides using a force field based on equivariant graph neural networks. The models demonstrate efficient learning from a small training dataset of energies, forces, stresses, and tensors of Born effective charges. We qualitatively capture the temperature dependence of the dielectric tensor and structural phase transitions in calcium titanate.

## Introduction

In atomistic simulations of realistic materials, one requires efficient models that can yield accurate properties, and at the same time are sufficiently fast to permit multiple sampling in order to acquire good statistics. At present, models based on graph convolutional neural networks (GCNNs) are some of the most promising choices to satisfy these requirements. These models have now been widely adopted for efficient modeling of properties of molecules and materials[1]. The accuracy of GCNNs can be further increased by replacing the convolution with a more general "message passing"[2], and imposing physics-informed equivariance constraints[3,4]. Equivariant message passing GNNs have been employed to implement interatomic potentials[5,6], predict spectra[7–9], reaction barriers[10], and atomic tensorial properties[11]. Their accuracy approaches that of ab initio methods, while locality ensures linear scaling with system size, making these methods ideal for large scale atomistic simulations. In addition to improving model accuracy, geometric equivariant features naturally support tensorial output values that represent various anisotropic properties. This makes equivariant GNNs well-suited for studying materials where directional dependencies and anisotropy play an important role.
In this work, message-passing equivariant GNNs are employed as a force field, as well as for predicting polarization from the tensors of Born effective charges, enabling the calculation of temperature-dependent anisotropic dielectric tensor from finite temperature MD simulations. The models are also accurate enough to be able to reproduce structural phase transitions with temperature, proving to be a powerful tool for probing thermodynamic properties and phase behavior of materials under realistic conditions. Here, the main focus is on the structure and properties of the mineral perovskite, $CaTiO_3$, but the approach used should be equally applicable to other systems. Equivariant GNNs exhibit fast training from small datasets, and facile preparation of accurate specialized models fitted to any target system of interest is possible. The models can then be used for long stable molecular dynamics runs in large cells, allowing calculations that would be unfeasible with cubically scaling density functional theory (DFT) approaches.

## Methods

Four datasets were employed for model training, labeled A, B, C, and D here. The letter suffix is also used for naming the model trained with the respective dataset. Dataset A contains energies, forces, stresses, and Born effective charges of 178 sparsified snapshot from ab initio MD simulations of Pnma $CaTiO_3$ between 300 K and 500 K. Dataset B contains 502 snapshots of $BaTiO_3$ at similar conditions, and dataset C 191 snapshots of $CaZrO_3$. Dataset D is a combination of datasets A-C. PBEsol functional was used throughout, and Hubbard $U$ correction of 3 eV was applied to $d$ electrons in transition metals, except for Ti, where $U = 0$ was used. MACE and Equivar models were trained from scratch using datasets A-D for 1000 epochs. An example of a learning curve of Equivar for Born effective charges is shown in Fig. S1, demonstrating fast learning, with a dataset of size ~100 being sufficient to have a MAE of $2 \times 10^{-2}$. For MACE, it is also found that a dataset of size ~100 is sufficient to exceed the accuracy of the foundation MACE-MP-0 model[12], underscoring the importance of the training set being close enough to the target system of interest.

Energies, forces, stress tensors, and tensors of atomic Born effective charges are obtained from MACE[6] and Equivar[11] models. The force $F_{\kappa,\alpha}$ acting on ion $\kappa$ along the direction $\alpha$ is given by:

$$F_{\kappa,\alpha} = F_{\kappa,\alpha}(\mathcal{E}=0) + |e|\Sigma_\beta \mathcal{E}_\beta Z^*_{\kappa,\beta\alpha} \quad (1)$$

Here, $\mathcal{E}_\beta$ is the applied external electric field, $|e|$ is the elementary charge, and $Z^*_{\kappa,\alpha\beta}$ is the tensor of Born effective charges. The forces at zero electric field $F_{\kappa,\alpha}(\mathcal{E}=0)$ are obtained from the MACE model, and Born effective charges from Equivar. The total energy is then given by:

$$E = E(\mathcal{E}=0) - |e|\Sigma_{\kappa,\alpha\beta} \mathcal{E}_\beta Z^*_{\kappa,\beta\alpha} u_{\kappa,\alpha} \quad (2)$$

with $u_{\kappa,\alpha}$ being the displacement of ion $\kappa$ along direction $\alpha$. The polarization is calculated from the finite field method, with change in polarization due to the displacement of ions being $P_\alpha = (e/\Omega)\Sigma_{\kappa,\beta} Z^*_{\kappa,\alpha\beta} u_{\kappa,\beta}$. Finite temperature MD simulations are performed with the external electric field $\mathcal{E} = 5 \times 10^7$ V/m.

## Results

Predictions of our MACE models are compared with those of the "foundation models"[12] trained on the Materials Project Trajectory (MPtrj) dataset[13]. We also compare the performance of the newly trained Equivar with the earlier BM1 model[11]. Of importance is the model's ability to capture the relevant physics of the system, in particular, to accurately reproduce the "anomalous" effective charges, which appear prominently in systems with mixed ionic-covalent bonding, such as perovskite oxides[14–16]. In perovskite titanates, the modulation of Ti $3d$ - O $2p$ orbital hybridization with interatomic distance[17,18] induces the flow of the electronic current along the Ti-O bonds, giving rise to anomalous charges[15,16]. Anomalous Born charges are a prerequisite for a large static dielectric constant[19,20], and are crucial for ferroelectricity[14,21]. The values for the anomalous charges in $ABO_3$ perovskites are ~7.1 for Ti, and ~-5.7 for O, whereas the "normal" charges are ~-2 for O and ~2.6 for Ca[22]. A satisfactory model must be able to predict the large anomalies in charges of Ti, vastly different charges for the two inequivalent O displacements, a mostly static charge of Ca, as well as capture the variations with atomic displacements.

Figure 1 compares model predictions with DFPT values for the evolution of the Born charges with atomic displacements in Pnma $CaTiO_3$ (CTO). The DFPT values that were used to train the models are in a good agreement with the earlier literature reports[22,23]. Note that the BM1 model[11], trained on a dataset of 1,224 perovskite, 17,991 $Li_3PO_4$, and 10,103 $ZrO_2$ structures, makes worse predictions than a specialized model with the same architecture (EquivarA) trained with only 178 $CaTiO_3$ structures. This is possibly due to the fact that no information on the $Z^*$ variation in CTO was available in the BM1 training set, whereas the dataset used for training EquivarA consisted of MD snapshots with perturbed geometries. Both anomalous and normal charges are well captured by EquivarA and EquivarD, whereas BM1 predictions sometimes deviate, e.g., by overestimating the decrease of the O charge with displacement. EquivarD, trained on a slightly more diverse chemistry, exhibits larger error than EquivarA, possibly due to regularization. The anomalous charges in CTO are seen to attain maximum values near the equilibrium positions.

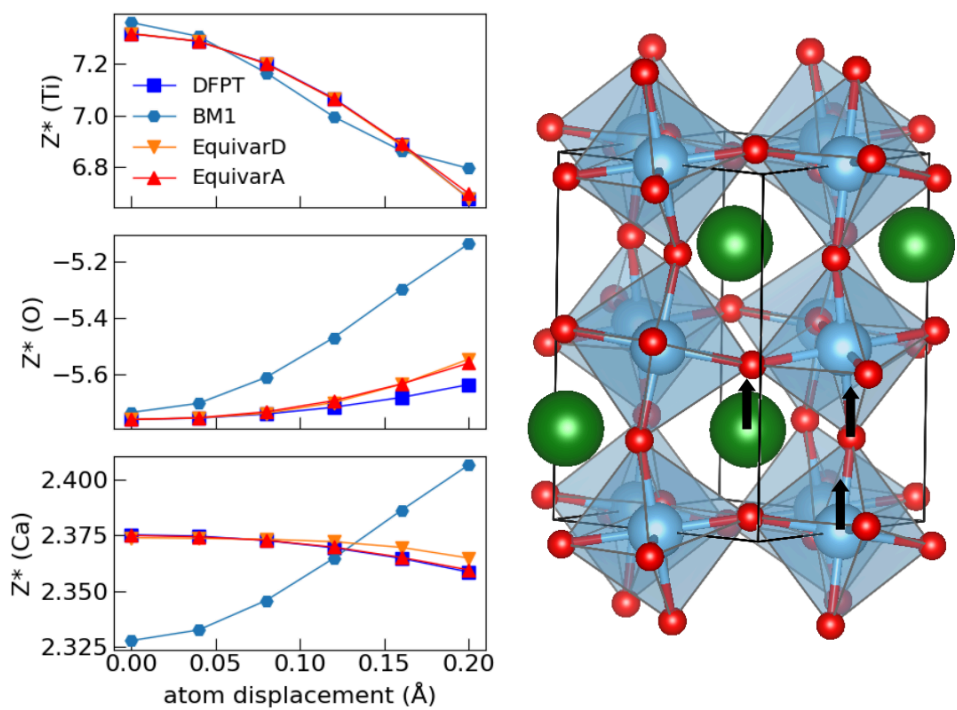

Figure 1. Dynamic charges for single atom displacements in $CaTiO_3$ computed with DFPT and different Equivar models. Displacements are in the direction of the *z* axis, as shown by the arrows in the panel on the right.

A range of custom MACE[6] force field models has been trained to perform MD simulations, and their performance compared with that of MACE-MP-0[12]. Figure 2a benchmarks force predictions of MACE-MP-0, as well as MACE models trained from scratch using the CTO dataset. In the latter case, a 5-fold cross-validation was carried out using the CTO dataset with 178 geometries. Despite the small dataset size, a specialized model significantly outperforms the generalist MACE-MP-0 trained with ~$10^6$ structures[13], with errors being more than an order of magnitude smaller. Similarly, we trained MACE models with a BTO and combined CTO+BTO+CZO datasets. Figure 2b compares the energy prediction for the soft phonon mode of $BaTiO_3$, again showing that specialized models give better predictions than MACE-MP-0. Note that MACE-MP-0 captures the energy profile near equilibrium quite well, but underestimates the energy away from equilibrium. The larger discrepancies for MACE-MP-0 may be partly due to the different functional being used (PBEsol here vs. PBE in MPtrj), as well as difference in values for the Hubbard $U$ correction. Still, the model performs surprisingly well for not being specifically trained with these configurations. Also note that the soft mode of BTO, for which long range electrostatic interactions play a great role, is captured well by MACE, despite being a local FF.

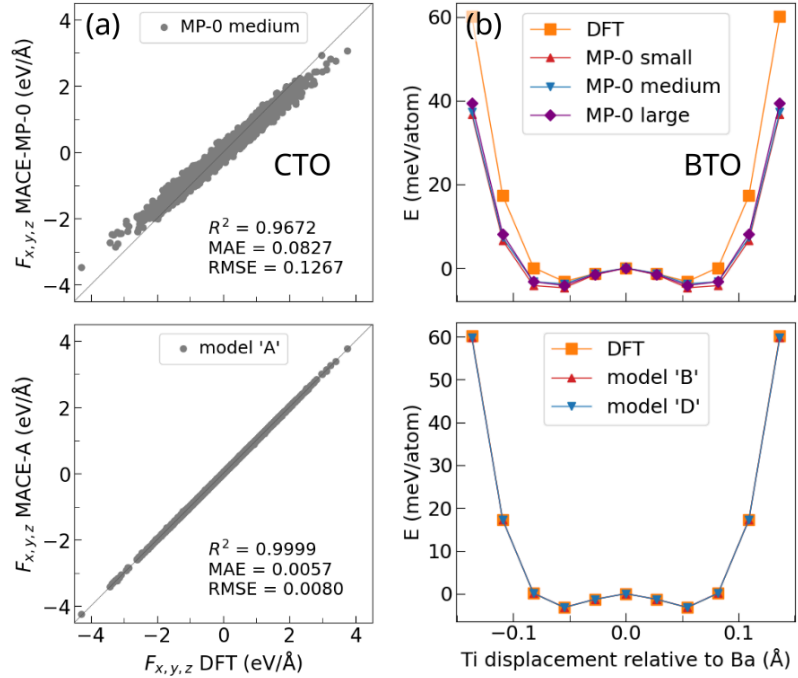

Figure 2. (a) Force parity plots for MACE-MP-0 (upper panel), and five-fold cross-validation of MACE model 'A' trained on the CTO dataset (lower panel) with PBESol DFT forces as a ground truth; (b) energy profiles for the displacement along the soft phonon mode in cubic $BaTiO_3$ using MACE-MP-0 (upper panel) and MACE models 'B' and 'D' (lower panel). PBESol DFT energies are also shown.

In order to obtain the polarization in the external field, a combined force field equal to the sum of MACE and Equivar was used, according to eqs 1-2. Total polarization is obtained from the Born charges and atomic displacements due to forces from the external field. Figure 3 shows the calculated static polarization of CTO for fields $0<E<10^8$ V/m. The softer MACE-MP-0 FF predicts larger polarization, and faster onset of nonlinearity compared to model 'A'. The predictions of the

$P$ values are seen to be more sensitive to the choice of the MACE force field than the Born charge model.

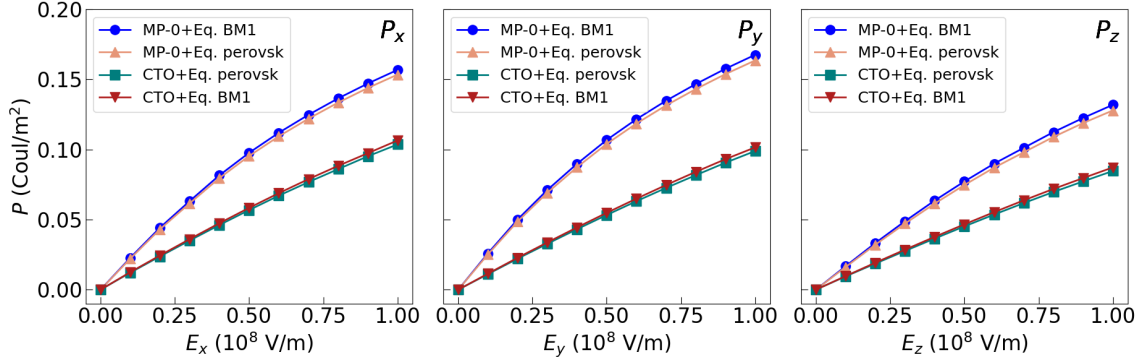

Figure 3. Polarization of CaTiO$_3$ in the applied electric field. The results for the MP-0 and custom trained MACE models are shown. From the slopes, the static dielectric tensor is obtained. MP-0 FF underestimates the force constants and overestimates the nonlinearity.

The efficient and accurate GCNN force fields enable fast simulations of large systems. In particular, statistical sampling for each atom, normally computationally prohibitive and restricted to model systems, becomes possible for realistic crystals. We test the capability of the force field to predict structural phase transitions in CTO. Experimentally, CTO undergoes two phase transitions, from orthorhombic to tetragonal, and from tetragonal to cubic structures, as temperature increases. The respective transition temperatures are 1512 ± 13 K and 1635 ± 2 K[24]. We performed *NpT* simulations using the 640-atom 4×4×2 and 1280-atom 4×4×4 supercells, and a heating rate of 1 K/ps, between 300 K and 1800 K.

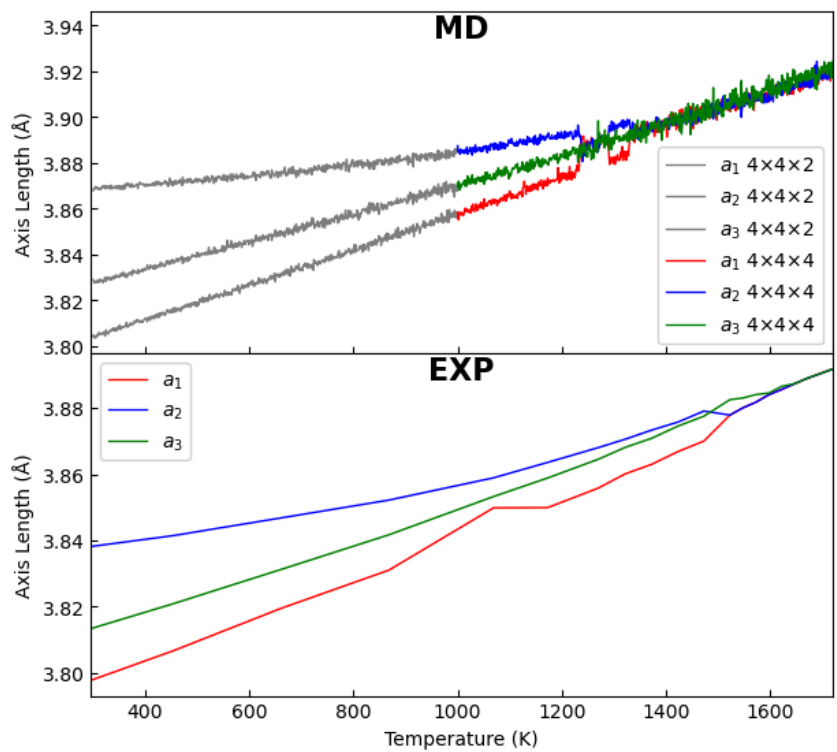

Figure 4. Structural phase transitions in CaTiO$_3$ from heating simulations and experimental data. Top panel shows the computed curves, and bottom panel experimental data[24]. In the upper panel, the gray lines show results for the 4×4×2 simulation cell, and color lines for the 4×4×4 cell.

The results of the heating simulations are compared with the experimental data in Fig. 4. Unlike experimental observations, a single phase transition from the orthorhombic to cubic phase occurs in the simulation, and no tetragonal phase is observed. During heating, the system switched from the orthorhombic to cubic phase and back, before adopting the symmetric cubic structure. The approximate temperature of the phase transition in the simulated system is around 1280 K. The system shows linear expansion throughout the range of temperatures tested, similar to experiment. Note that the structural phase transitions in CTO are rather subtle, and different transition temperatures as well as the presence of another orthorhombic phase have been reported in previous experiments, as summarized in Table I of Ref.[25]. The lower predicted phase transition temperature and the absence of the tetragonal phase could also be due to error inherent in the approximate density functional used for the training set creation, and not indicative of the model accuracy.

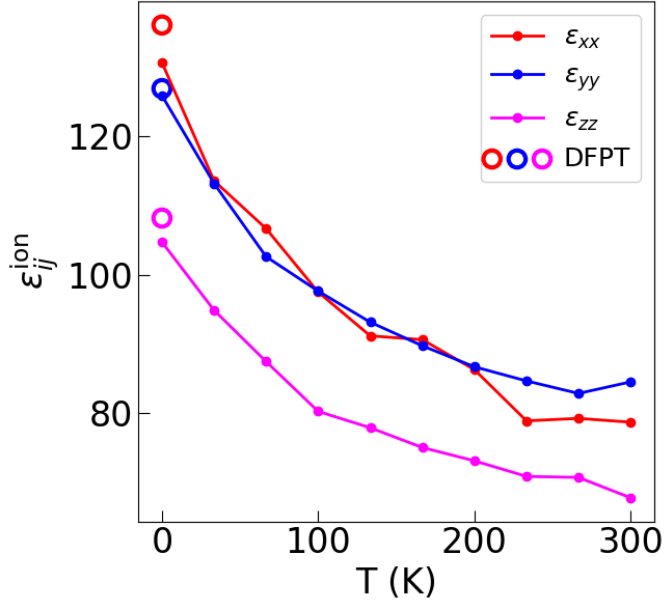

Figure 5. Temperature-dependent dielectric tensor of $CaTiO_3$. The values of the ionic part of the dielectric tensor calculated using finite temperature MD are shown with filled circles, while zero temperature DFPT values are shown with empty circles.

Finally, we demonstrate finite temperature molecular dynamics simulations to evaluate the dielectric tensor of CTO. Experimentally, it is known to decrease with increasing temperature, depicted by a characteristic bell-shaped curve[26]. We performed MD simulations of CTO at finite $T$ and recorded a full dielectric tensor by averaging the polarization at each temperature for each direction of the electric field parallel to the axes of the Cartesian coordinate system. The resulting diagonal components are shown in Fig. 5. The simulations are seen to reproduce the decreasing trend and also the shape of the experimental curve. Note that the theoretical values are smaller than experimental ones due to the limitations of density functionals used to generate the ground truth set. Also, the low temperature plateau stemming from the quantum mechanical zero point motion, is absent. This effect cannot be captured in classical simulations and requires a quantum mechanical description of the nuclear motion.

## Conclusions

We demonstrate the application of force fields based on equivariant graph neural networks to finite temperature MD simulations of realistic materials in external electric fields. MACE and Equivar models trained on small datasets of perovskite oxides show better performance than the generalist models trained on much larger sets. Using these models, total energies, forces and tensors of dynamic Born effective charges are obtained with close to ab initio accuracy, and molecular dynamics simulations at finite temperatures and in finite electric fields are performed. The decrease of the dielectric tensor with temperature and structural phase transition from orthorhombic to cubic phase are captured qualitatively in calcium titanate.